\documentclass[twocolumn,english,prb]{revtex4-1}

\usepackage[utf8]{inputenc}
\usepackage[T1]{fontenc}
\usepackage{graphicx}
\usepackage{amsmath}
\usepackage{color}

\DeclareGraphicsExtensions{.pdf,.eps,.mps,.png,.jpg}

\begin{document}

\title{Pauli blockade microscopy of quantum dots}

\author{E. Wach}
\affiliation{AGH University of Science and Technology, Faculty of Physics and
Applied Computer Science,\\
 al. Mickiewicza 30, 30-059 Krak\'ow, Poland}

\author{B. Szafran}
\affiliation{AGH University of Science and Technology, Faculty of Physics and
Applied Computer Science,\\
 al. Mickiewicza 30, 30-059 Krak\'ow, Poland}

\begin{abstract}
We propose a spin-sensitive scanning probe microscopy experiment on double quantum dots in Pauli blockade conditions.
Electric spin resonance is induced by an AC voltage applied to the scanning gate which induces lifting of the Pauli blockade
of the current.  The stationary Hamiltonian eigenstates are used as a basis for description of the spin dynamics with the AC potential of the probe. For the two-electron system we evaluate the transitions rates from triplet $\mathrm{T}_+$ state to singlet $\mathrm{S}$ or triplet $\mathrm{T}_0$ states, i.e. to conditions in which the Pauli blockade of the current is lifted. The rates of the spin-flip transitions are consistent with the transition matrix elements and strongly dependent on the tip position. Probing the spin densities and identification of the final transition state are discussed.
\end{abstract}
 
\maketitle

\section{Introduction}

The potential landscape within the electron gas buried shallow
beneath the surface of semiconductor  \cite{qpc0,qpc1,qpc8,qpc10,sqd1,sqd2,mqd1,wff,sekolas}, quantum wires  \cite{mqd2,boyde}, carbon nanotubes \cite{mqd3,zhou,ultra},  graphene \cite{connolly,Morikawa2015,Bhandari2016}
can be modified in a controllable manner by potential
of an external gate. This fact was 
used for development of a scanning gate microscopy technique  \cite{sgmr,scar},
in which the current or conductance maps are taken
as a function of the position of a charged tip of an atomic
force microscope floating above the surface of the system  \cite{sgmr,scar}.
The technique produces spatial information on the charge
transport in open systems, including quantum point contacts  \cite{qpc0,qpc1,qpc8,qpc10}, or quantum rings \cite{sgmr}, magnetic focusing  \cite{wff,Morikawa2015,Bhandari2016}, etc. 
The method was also used to probe the states localized in quantum dots \cite{sqd1,sqd2,mqd1,mqd2,boyde,mqd3,ultra,connolly}, where the potential of the tip switches on
or off the Coulomb blockade  \cite{cb} of the current flow. The technique was nicknamed a Coulomb blockade microscopy \cite{qiang,boyd,wachjpcm,wachssc}.

\begin{figure}[h!t]
\begin{center}
\includegraphics[width=0.48\textwidth]{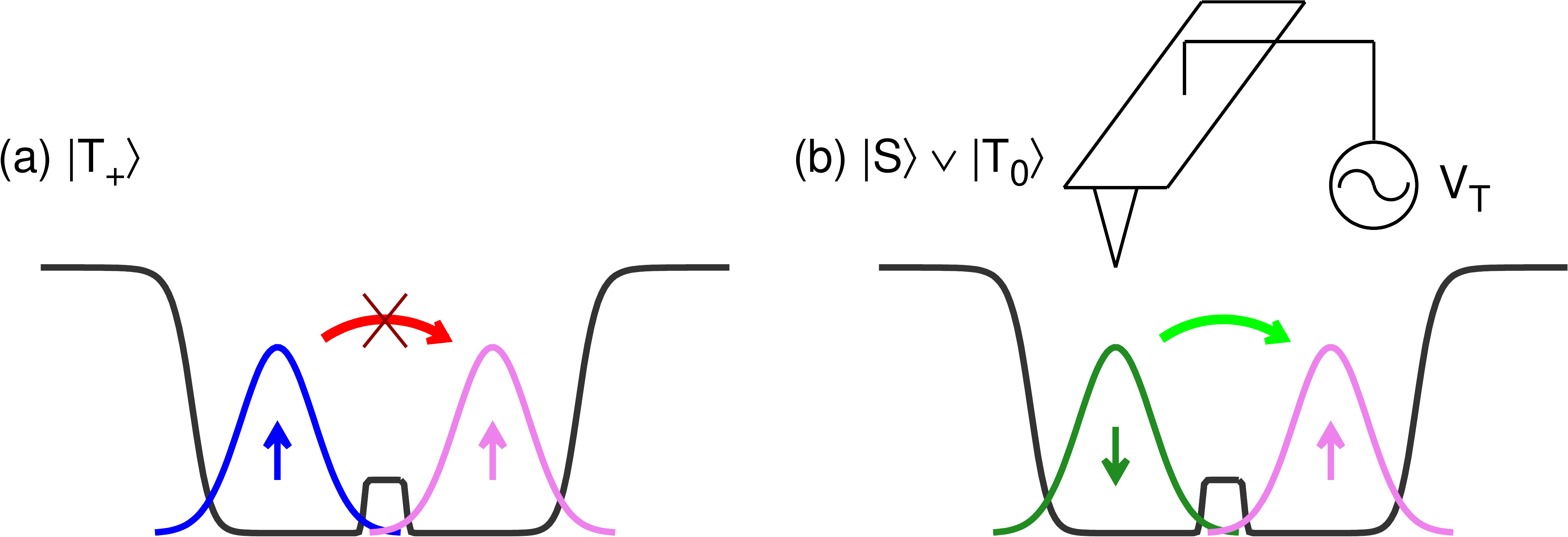}
\caption{Schematic drawing of the system: the confining potential $V_\mathrm{c}(x, 0)$ of a double quantum dot, each dot stores a~single electron. (a)~The system is prepared in the spin-polarized triplet state $|\mathrm{T}_+\rangle$. Both spins are identical (up), the electron hopping is forbidden. (b)~The~AC voltage is applied to the AFM tip and causes the spin-flip; the Pauli blockade is lifted.
}
\label{model}
\end{center}
\end{figure}

In the present paper we propose to use the external gate of the atomic force microscope on
the system in which the flow of the current is blocked by the Pauli exclusion principle.
The most elementary system in which the Pauli blockade \cite{pauliblockade,pauliblockade1,koppens,extreme,stroer,nadji,laird} is the double quantum dot
biased in a manner that each of the dots stores a single electron and the electrons possess identical spins [see Fig. 1(a)].
The electron hopping from one dot to the other can occur only provided that the spin of one of the electrons is flipped.
The spin-flip can be intentionally induced by electron spin resonance using microwave radiation \cite{koppens} or, 
in spin-orbit coupled systems, by electric-dipole spin resonance driven by the AC voltage applied 
to one of the gates \cite{nadji,extreme,stroer,edsr,laird,weg}. 
The idea investigated in this paper is to use the scanning probe as the gate which provides the AC perturbation [see Fig. \ref{model}]
that drives the spin polarized triplet $\mathrm{T}_+$ to systems in which the Pauli blockade is lifted: the singlet $\mathrm{S}$ or $\mathrm{T}_0$,
which in III-V materials are mixed by the hyperfine field \cite{mixed}.
The maps of the transition rates bear signatures of the single-electron wave functions or the convolution
of the majority and minority spin components with the potential of the tip. We demonstrate that it is possible to 
perform imaging of the spatial distribution of the spin in a confined spin-orbit coupled system and to
identify the 
final state $\mathrm{S}$ or $\mathrm{T}_0$ of the spin-transition lifting the Pauli blockade by the form of the map.

\section{Theory}

\subsection{Solution of the time-independent Schr\"odinger equation}

We investigate a single electron confined in a rectangular quantum dot and an electron pair in two tunnel-coupled quantum dots assuming a strictly two-dimensional model of confinement. The one-electron Hamiltonian
\begin{equation}
 h =  \left(\frac{ \left( -i \hbar \mathbf{\nabla} + e\mathbf{A} \right)^2}{2m^*} + V_\mathrm{c}(\mathbf{r}) \right) \mathbf{1} + H_{\mathrm{SO}} + \frac{1}{2}g^*\mu_\mathrm{B}B\sigma_z 
 \label{h_1}
\end{equation}
is taken in the effective-mass-approximation form, where $m^*$ and $g^*$ stand for the effective mass and Land\'{e} factor, respectively, $\mathbf{1}$ is the identity matrix, $V_\mathrm{c}(\mathbf{r})$ is the confinement potential, and $H_{\mathrm{SO}}$ introduces Dresselhaus spin-orbit interaction~\cite{Dresselhaus}. The last term describes the spin Zeeman effect in the presence of a static magnetic field $B$ oriented perpendicular to the plane of confinement. The symmetric gauge $\mathbf{A} = B \left( -\frac{y}{2}, \frac{x}{2}, 0 \right)$ is used.

We account for the linear Dresselhaus SO term 
\begin{equation}
 H_{\mathrm{SO}}=\gamma\left< k^2_z \right> \left[ \sigma_x k_x - \sigma_y k_y \right].
\end{equation}
The cubic component of the Dresselhaus term  produces small effects as compared to the linear one and thus is neglected~\cite{small_cubic}. $\sigma_x, \sigma_y, \sigma_z$ are Pauli matrices and $k_x, k_y, k_z$ are the components of the wave vector $\mathbf{k}$, while $\hbar \mathbf{k} = \mathbf{p} = -i \hbar \mathbf{\nabla} + e\mathbf{A}$. 
The linear Dresselhaus constant $\gamma\left< k^2_z \right>$ for two-dimensional Hamiltonian is obtained by 
\begin{equation}
 \gamma\left< k^2_z \right>=\gamma\frac{\pi^2}{d^2},
\end{equation}
where $\gamma$ is the Dresselhaus bulk SO coupling constant and $d$ means the height of the QD ($d=5.42~\mathrm{nm}$ is taken). We use $\gamma=32.2~\mathrm{meV nm}^3$ assuming the QD is made of In$_{0.5}$Ga$_{0.5}$As alloy \cite{gamma}, for which $\gamma\left<k^2_z\right>~\approx~10.82~\mathrm{meV nm}$. The other parameters are calculated as arithmetic average of GaAs and InAs values, i.e. $m^*=0.0465 m_0$, $g^*=-8.97$ \cite{mass, g_factor} and dielectric constant $\epsilon=13.5$ \cite{Winkler_SO}.  We assume that the 2DEG is symmetrically doped. The potentials inducing the lateral confinement in quantum dots are shallow and the corresponding electric fields are weak which allows us to neglect the Rashba spin-orbit interaction.

\begin{figure}[h!t]
\begin{center}
\includegraphics[width=0.48\textwidth]{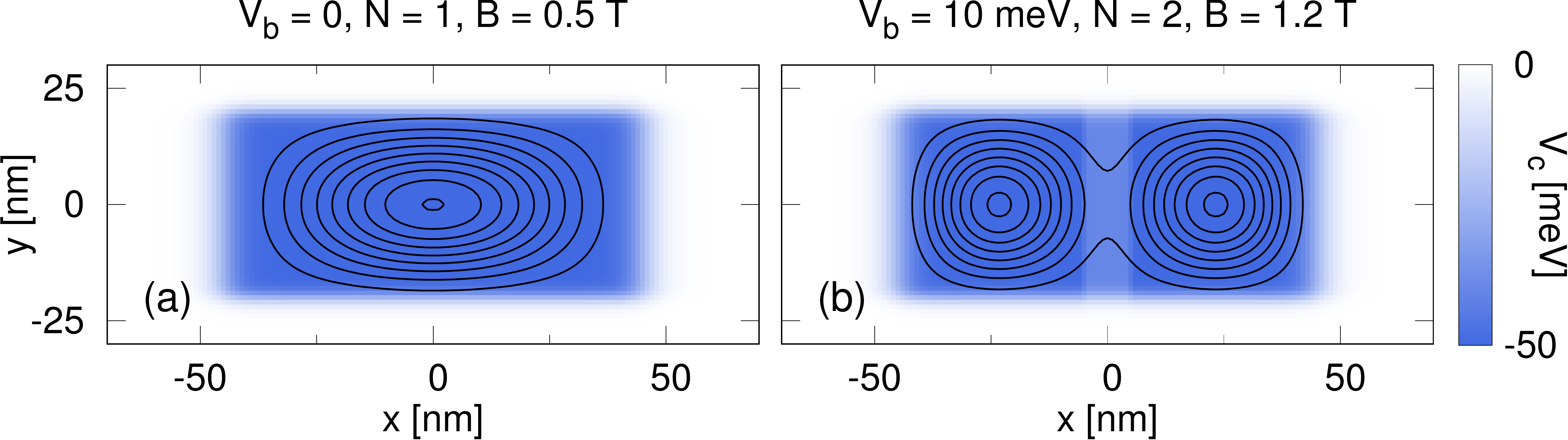}
\caption{The contours of the charge density for (a)~a single electron ($N=1$) confined in the single rectangular elongated quantum dot at $B=0.5~\mathrm{T}$; (b)~two electrons ($N=2$) in two tunnel-coupled quantum dots at $B=1.2~\mathrm{T}$. The shades of blue show the confining potential $V_\mathrm{c}(x, y)$.
}
\label{charge}
\end{center}
\end{figure}

In the following we consider confinement potential of the profile
\begin{eqnarray}
&& V_\mathrm{c}(\mathbf{r})=-\frac{V_0}{\left( 1+\left[ \frac{x^2}{R_x^2} \right]^{10} \right) \left( 1+\left[ \frac{y^2}{R_y^2} \right]^{10} \right) } \nonumber \\ &&+ \frac{V_\mathrm{b}}{\left( 1+\left[ \frac{x^2}{R_\mathrm{b}^2} \right]^{10} \right) \left( 1+\left[ \frac{y^2}{R_y^2} \right]^{10} \right) },
\end{eqnarray}
with the depth of the quantum dot/dots $V_0=50~\mathrm{meV}$ and a flat bottom of nearly rectangular shape with dimensions $2R_x \times 2R_y$ ($R_x=45~\mathrm{nm}, R_y=20~\mathrm{nm}$). $R_\mathrm{b}=5~\mathrm{nm}$ defines the interdot barrier thickness ($2R_\mathrm{b}$) and $V_\mathrm{b}$ is the barrier height. In the case of a single quantum dot $V_\mathrm{b}=0$. For two tunnel-coupled quantum dots we assume $V_\mathrm{b}=10~\mathrm{meV}$. The confining potential and the ground-state charge density for both the single and the double quantum dot are shown in Fig. \ref{charge}.

We solve the one-electron eigenequation by diagonalization using the basis of multicenter Gaussian functions $f_p(\mathbf{r})$:
\begin{equation}
 f_p(\mathbf{r}) = \frac{1}{\alpha_\mathrm{G} \sqrt{\pi}} \exp \left[ -\frac{ \left( \mathbf{r}-\mathbf{R}_p \right)^2}{2 \alpha_\mathrm{G}^2} +\frac{ieB}{2 \hbar} \left( xY_p - yX_p \right) \right].
\end{equation}
The centers of Gaussians $\mathbf{R}_p = (X_p, Y_p)$ are distributed on a rectangular mesh of $M \times N$ points. We use $25 \times 11$ basis functions. The $n$-th eigenfunction is expanded as
\begin{equation}
 \psi_n(\mathbf{r}, s) = \sum_{p}^{M \times N} \sum_{s\in \{\uparrow, \downarrow\}}  a_{p, s}^{(n)} \chi_s f_p(\mathbf{r}),
\end{equation}
where $\chi_s$ are the eigenstates of $\sigma_z$ matrix. The parameter $\alpha_\mathrm{G}$ and the distance between the centers $\mathbf{R}_p$ are optimized variationally.

The eigenequation of two-electron Hamiltonian
\begin{equation}
 H = h_1 + h_2 + \frac{e^2}{4 \pi \epsilon \epsilon_0 |\mathbf{r}_{1} - \mathbf{r}_{2}|}
\end{equation}
is solved by the configuration-interaction method. Two-electron wave functions are expressed by the linear combinations of the Slater determinants formed from the spin-orbitals $\psi_k(\mathbf{r}, s)$:
\begin{eqnarray}
 \Psi_n(\mathbf{r}_1, \mathbf{r}_2, s_1, s_2) & = & \frac{1}{\sqrt{2}} \sum_k b_k^{(n)} \left[ \psi_{k_1}(\mathbf{r}_1, s_1)\psi_{k_2}(\mathbf{r}_2, s_2) \right. \nonumber \\ 
 ~ & ~ & \left. - \psi_{k_2}(\mathbf{r}_1, s_1)\psi_{k_1}(\mathbf{r}_2, s_2) \right].
\end{eqnarray}
We take $30$ single-electron spin-orbitals which give convergence of the energies with a precision better than $0.1~\mathrm{\mu eV}$ and produce $\binom{30}{2} = 435$ Slater determinants.



We assume the effective tip potential is short-range~\cite{SZ_ring} and use the Lorentzian formula
\begin{equation}
 V_\mathrm{T}\left( \mathbf{r}; \mathbf{r}_\mathrm{tip} \right) = \frac{V_\mathrm{tip}}{\left(\frac{x-x_\mathrm{tip}}{d_\mathrm{tip}}\right)^2 + \left(\frac{y-y_\mathrm{tip}}{d_\mathrm{tip}}\right)^2 + 1},
 \label{V_T}
\end{equation}
where $V_\mathrm{tip}=5~\mathrm{meV}$ is the height of the potential maximum induced by the tip placed at position $\mathbf{r}_\mathrm{tip} = \left( x_\mathrm{tip}, y_\mathrm{tip}\right)$ above the confinement plane. $d_\mathrm{tip}=20~\mathrm{nm}$  defines the width of the Lorentzian. 
The time dependence is introduced to Hamiltonian by the periodic modulation of the tip potential:
\begin{equation}
 h'(t) = h + V_\mathrm{T}\left( \mathbf{r}; \mathbf{r}_\mathrm{tip} \right) \sin(\omega t) \label{odczasu}
\end{equation}
for the one-electron Hamiltonian, and
\begin{equation}
 H'(t) = H + \big[  V_\mathrm{T}\left( \mathbf{r}_1; \mathbf{r}_\mathrm{tip} \right)  + V_\mathrm{T}\left( \mathbf{r}_2; \mathbf{r}_\mathrm{tip} \right) \big] \sin(\omega t)
\end{equation}
for the two-electron case.

For the single electron the system is prepared in an initial state $\psi_i(\mathbf{r}, s) = \psi_0(\mathbf{r}, s) \equiv \psi(\mathbf{r}, s, t=0)$. The expansion in the basis given by the eigenstates $\psi_n(\mathbf{r}, s)$
\begin{equation}
 \psi(\mathbf{r}, s, t) = \sum_n c_n(t) \psi_n(\mathbf{r}, s) \exp\left( -\frac{iE_n t}{\hbar} \right)
 \label{psi_expansion}
\end{equation}
describes time evolution of the wave function.
By inserting the expansion (\ref{psi_expansion}) to the time-dependent Schr\"odinger equation
\begin{equation}
 i \hbar \frac{\partial \psi(\mathbf{r}, s, t)}{\partial t} = h'(t) \psi(\mathbf{r}, s, t)
\end{equation}
one can obtain a system of differential equations
\begin{eqnarray}
  \frac{dc_k(t)}{dt} & = & - \frac{i}{\hbar} \sum_n c_n(t) \sin(\omega t) \cdot \nonumber \\ 
  && \cdot \langle \psi_k | V_\mathrm{T}\left( \mathbf{r}; \mathbf{r}_\mathrm{tip} \right) | \psi_n \rangle \exp \left( -i\omega_{nk}t \right),
  \label{eq_for_ck}
\end{eqnarray}
where $\omega_{nk} = \frac{E_n-E_k}{\hbar}$.
The equations are solved for the expansion coefficients $c_k(t)$ using finite-difference methods: iterative Crank-Nicolson scheme and the Askar-Cakmak method~\cite{Askar}.

Finally, the probability of effecting a transition from state $\psi_i(\mathbf{r}, s)$ to state $\psi_k(\mathbf{r}, s)$ after time $t$ is given by
\begin{equation}
 P_{i\rightarrow k} (t) = | \langle \psi_k | \psi(t) \rangle |^2 = |c_k(t)|^2.
\end{equation}

 The two-electron time-dependent Schr\"odinger equation for $\Psi(\mathbf{r}_1, \mathbf{r}_2, s_1, s_2, t)$ is solved in the same way.

\subsection{Time-dependent perturbation theory}

The exact solution of Eq. (\ref{eq_for_ck}) produces the direct (single-photon) and the fractional (multiphoton) resonances for the frequencies $\omega=\omega_{ki}$ and $\omega=\omega_{ki}/m$ ($m\in \{2, 3, \hdots\}$), respectively. The direct resonances, that are distinctly wider then the fractional ones, can be explained within the time-dependent perturbation theory of the first order. In the perturbation expansion 
\begin{equation}
 c_k(t) = c_k^{(0)} + c_k^{(1)}(t) + \hdots 
\end{equation}
$c_k^{(0)}$ is independent of time and corresponds to the initial state: $c_k^{(0)} = \delta_{ki}$.
The formula for the first-order correction reads as
\begin{equation}
 c_k^{(1)}(t) = -\frac{i}{\hbar} \int_{0}^{t} \langle \psi_k |V_\mathrm{T}| \psi_i  \rangle  \sin\left( \omega t'\right) \exp\left( i\omega_{ki} t'\right) dt'. \label{ford}
\end{equation}
In particular, for $k\ne i$ the probability $P_{i\rightarrow k} (t)$ in the first-order approximation depends mainly on the matrix elements $\langle \psi_k |V_\mathrm{T}| \psi_i \rangle$.

\section{Results and Discussion}

This Section is organized in the following way. First we discuss the spin-flipping transition
for a single electron in a single quantum dot [Fig. \ref{charge}(a)]. Then, we proceed to description of spin transitions in the two-electron system for a double quantum dot [Fig. \ref{charge}(b)]. 

\subsection{Single electron in elongated quantum dot}
\begin{figure}[h!t]
\begin{center}
\includegraphics[width=0.4\textwidth]{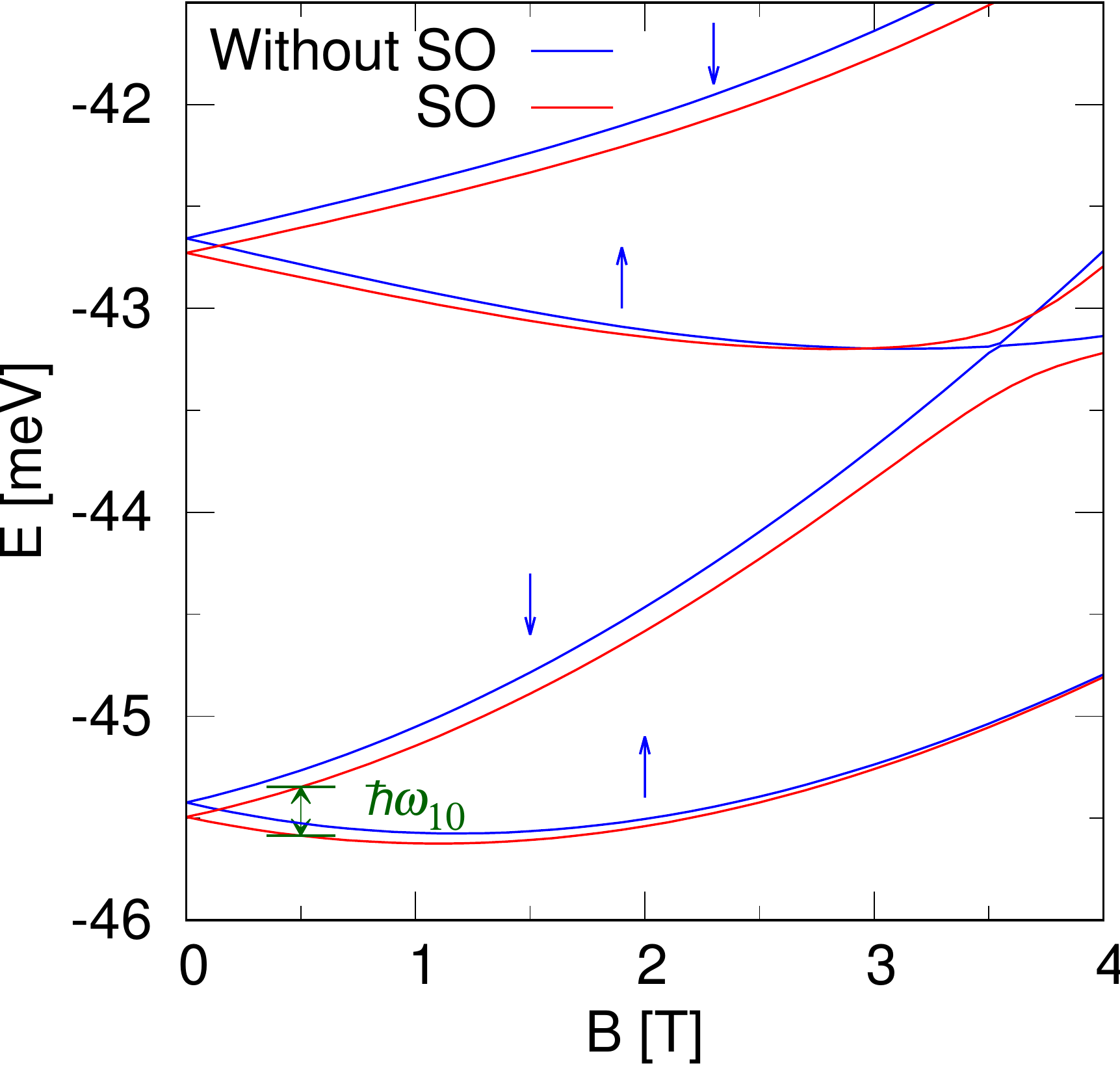}
\caption{The lowest single-electron energy levels as functions of the magnetic field for the rectangular elongated quantum dot ($V_\mathrm{b}=0$). Blue (red) lines show the levels obtained without (with) spin-orbit coupling. Blue arrows indicate the spin directions for corresponding energy levels without SO coupling. Green arrows mark the resonant frequency in the case of SO coupling for $B=0.5~\mathrm{T}$, $\omega_{10}=\frac{E_1-E_0}{\hbar}$.
}
\label{energy_QD1}
\end{center}
\end{figure}

The single-electron energy spectrum for the quantum dot of Fig. \ref{charge}(a) is displayed
in Fig. \ref{energy_QD1}. The energy effect of the spin-orbit coupling is small and amounts
in a shift of the energy levels for $B=0$, and opening of the avoided crossings between  energy 
levels that possess opposite spins in the absence of SO coupling. 

\begin{figure}[h!t]
\begin{center}
\includegraphics[width=0.4\textwidth]{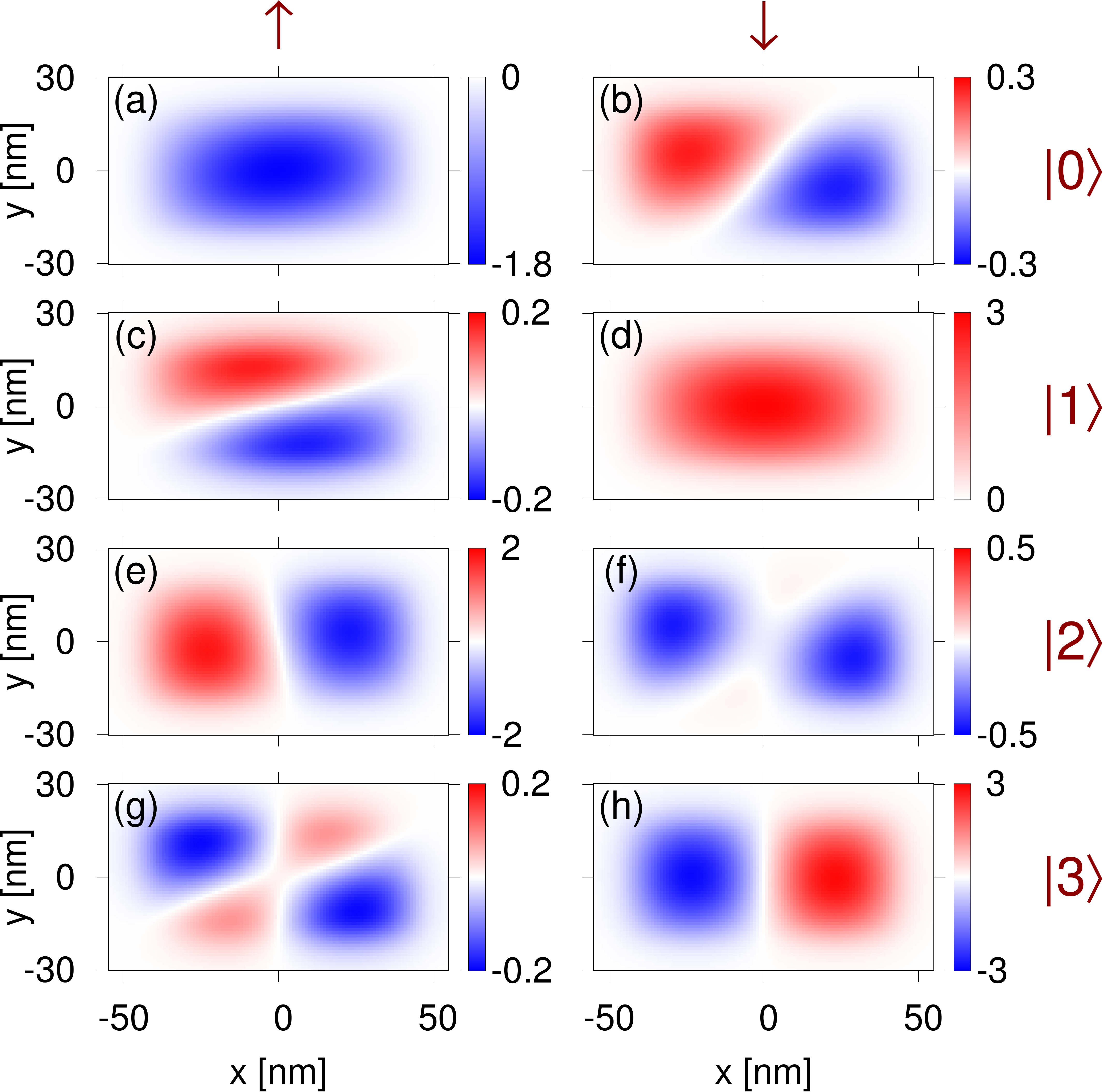}
\caption{Spin up (left column) and spin down (right column) components of single-electron wave functions for $B=0.5~\mathrm{T}$. The color scales give the real part of wave functions in units of $[10^{-2}~\mathrm{nm}^{-1}]$. The ground state is denoted as $|0\rangle \equiv |\psi_0\rangle$ (cf. first row of plots), first excited state as $|1\rangle$, etc.
}
\label{wf1e_QD1}
\end{center}
\end{figure}

The ground-state and the first excited state at $B=0$ remain
two-fold degenerate in presence of the SO coupling. The states of the Kramers doublets correspond to opposite eigenvalues of $P\sigma_z$ operator \cite{spar}, where $P$ stands for the parity operator $Pf(r)=f(-r)$.
The components of the wave functions are given in Fig. \ref{wf1e_QD1}. The majority and minority spin components possess 
opposite symmetries with respect to point inversion 
across the origin. 

\begin{figure}[h!t]
\begin{center}
\includegraphics[width=0.4\textwidth]{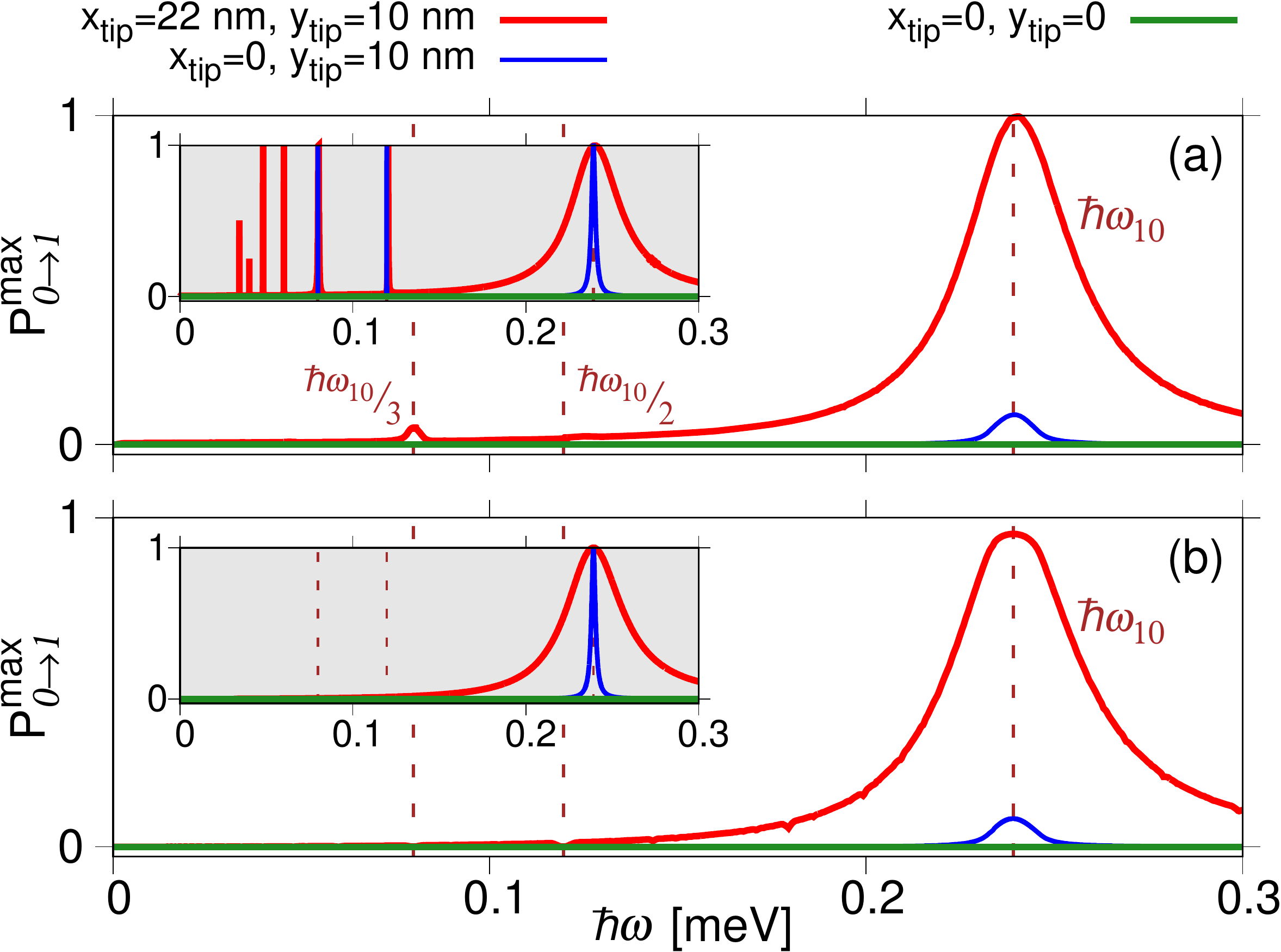}
\caption{Maximal probability of finding the electron in $|1\rangle$ state as a function of the AC frequency for three positions of the tip that are defined in a color legend. Vertical dashed lines indicate a few resonant frequencies: $\hbar\omega_{10}$, $\hbar\omega_{10}/2$, $\hdots$. (a)~Results of the solution of the time-dependent Schr\"odinger equation for a single electron. (b)~Results obtained with the time-dependent perturbation theory of the first order. The simulations with the initial state $|\psi_i\rangle = |0\rangle$ lasted $300~\mathrm{ps}$ ($250~\mathrm{ns}$ for the insets).
}
\label{prob10_QD1}
\end{center}
\end{figure}

We now consider the spin-flip or the transition between the 
ground state energy level and the first-excited energy level 
separated by $\hbar \omega_{10}$ [see Fig. \ref{energy_QD1}]. 
For the initial condition of the spin-flip process we set the system in the ground-state,
then we turn on the AC perturbation by the tip as given by Eq. (\ref{odczasu}). The maximal occupation of the spin-down
state as calculated by the exact solution of the Schr\"odinger equation is plotted in Fig. \ref{prob10_QD1}(a) for varied tip locations and the AC pulse of $300$~ps. For the tip at the center of the QD potential no spin transitions are observed (green line). When the tip is moved slightly off the origin (blue line) the transition becomes visible. For the tip off the symmetry axis of the QD potential (red line) the transition occurs much faster. 
The inset to Fig.~\ref{prob10_QD1}(a) shows the results for the duration of the AC signal increased to $250$~ns. The transition probability reaches 1 for the tip off the origin, only the width of the transitions depends on the tip position. Besides the first order transition near $\hbar\omega=\hbar\omega_{10}$ the
fractional resonances corresponding to multiphoton transitions 
are seen at lower energy side.  

\begin{figure}[h!t]
\begin{center}
\includegraphics[width=0.4\textwidth]{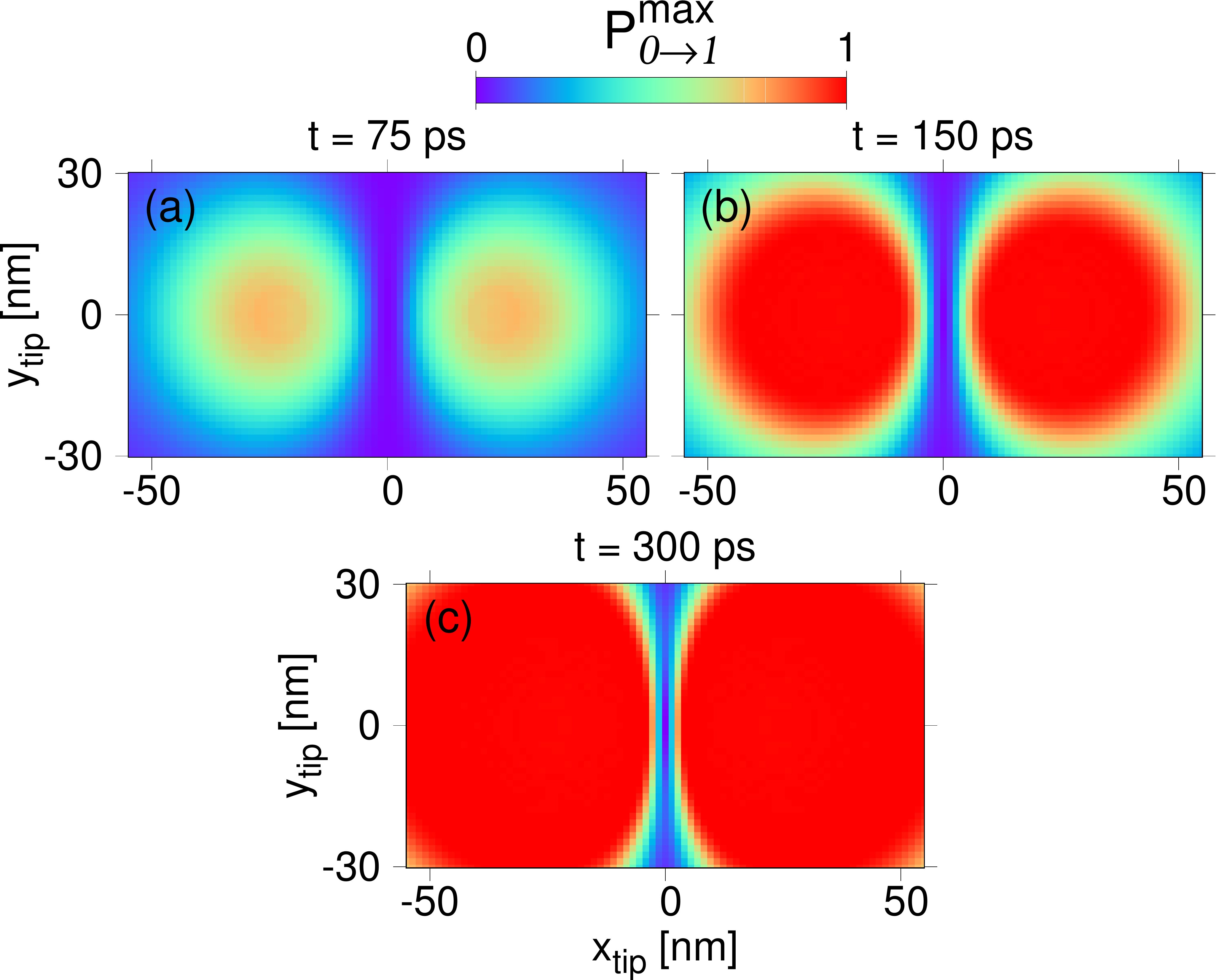}
\caption{Maximal occupation of the first excited state $|1\rangle$ for simulation lasting (a)~$75~\mathrm{ps}$, (b)~$150~\mathrm{ps}$, (c)~$300~\mathrm{ps}$ as function of the tip position. The frequency $\hbar\omega = 0.2392~\mathrm{meV}$ used in the calculations corresponds to the direct resonance (see Fig.~\ref{prob10_QD1}(a)).
}
\label{map10_QD1}
\end{center}
\end{figure}

\begin{figure}[h!t]
\begin{center}
\includegraphics[width=0.4\textwidth]{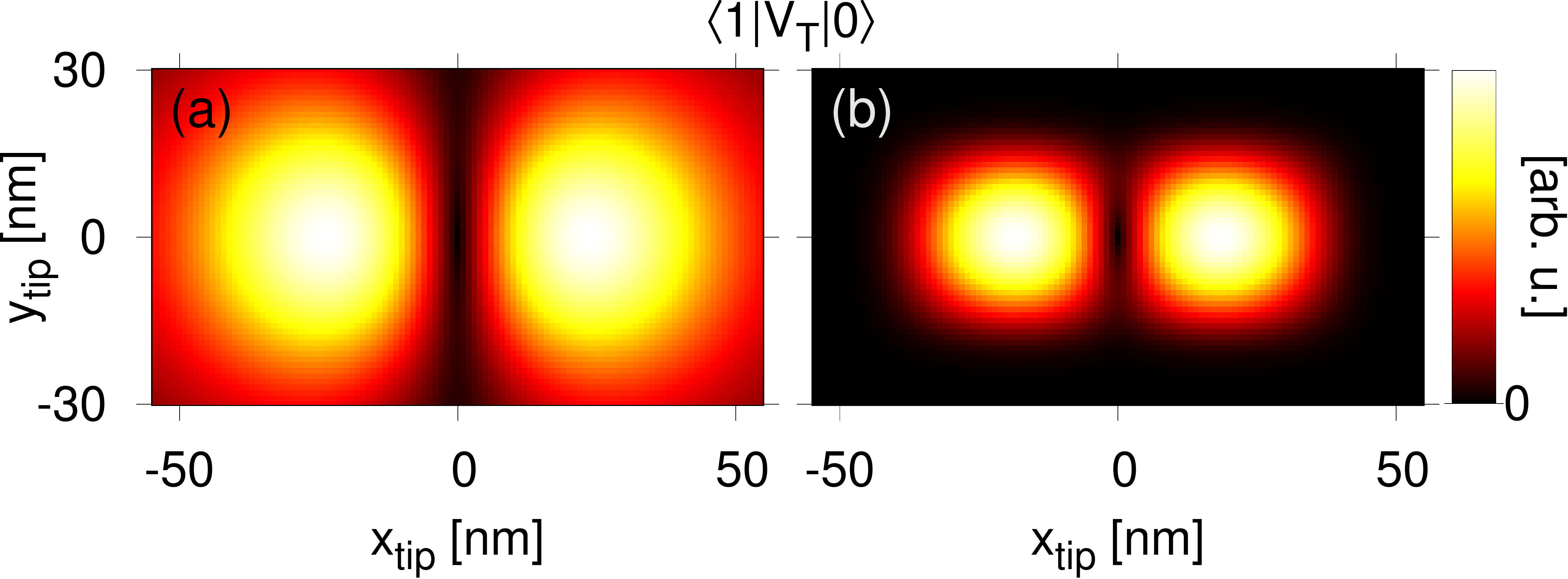}
\caption{(a) The absolute value of $\langle 1|V_\mathrm{T}|0 \rangle$ matrix element; panel (b) shows its integrand for $x=x_{\mathrm{tip}}, y=y_{\mathrm{tip}}$. 
}
\label{matrix_el_QD1}
\end{center}
\end{figure}

The transition probability as a function of the tip position 
for $\hbar\omega=\hbar\omega_{10}$ is displayed in Fig. \ref{map10_QD1}. In order to understand the dependence we
use the first order perturbation theory, which correctly
reproduces the direct spin-flipping transition -- see Fig. \ref{prob10_QD1}(b). The transition to the excited state 
according to the first order perturbation [Eq. (\ref{ford})] depends on the matrix element $\langle 1 | V_\mathrm{T} | 0 \rangle$ 
which is plotted in Fig. \ref{matrix_el_QD1}(a). The transition
matrix element is strictly zero for $x_\mathrm{tip}=y_\mathrm{tip}=0$ since
the integrands for the spin-up and spin-down components [Fig. \ref{wf1e_QD1}(a-d)] are of opposite parity. 
For that reason the transition matrix element for the tip at the symmetry center of the system vanishes. 
The matrix 
element possesses maxima off the $x_\mathrm{tip}=0$ axis of the plot [Fig. \ref{matrix_el_QD1}(a)] in consistence with 
the results of the time-dependent simulation [Fig. \ref{map10_QD1}(a-c)]. Concluding,
the spin-flip probability maps as functions of the tip position extract the form of the matrix elements
with the integrands corresponding to the convolution of the minority/majority spin components of the wave functions
with the potential of the tip. For the latter in the delta-like limit the matrix element tends
to the overlap of the minority/majority spin components of the wave function. The form
of the integrand of the matrix element for the adopted  tip width of $d_\mathrm{tip}=20$~nm is displayed in Fig. \ref{matrix_el_QD1}(b).

\subsection{Double quantum dot}

\begin{figure}[h!t]
\begin{center}
\includegraphics[width=0.4\textwidth]{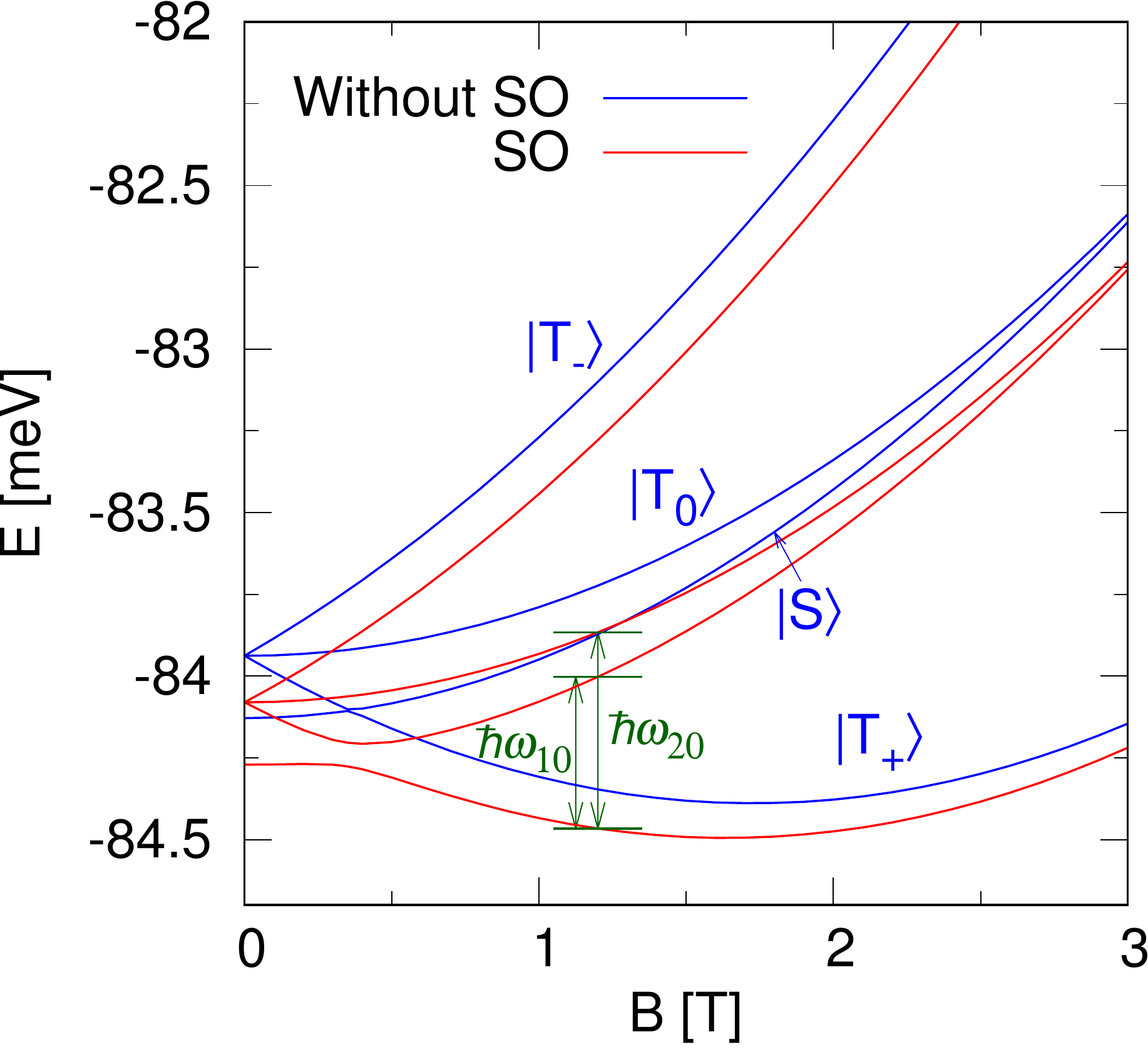}
\caption{The lowest two-electron energy levels as function of the magnetic field for two tunnel-coupled quantum dots ($V_\mathrm{b}=10~\mathrm{meV}$). Blue (red) lines show the levels obtained without (with) spin-orbit coupling. Blue labels $|\mathrm{S}\rangle$ and $|\mathrm{T}_+\rangle$, $|\mathrm{T}_0\rangle$, $|\mathrm{T}_-\rangle$ indicate the energy levels of a singlet and the triplet states without SO coupling. Green arrows mark the resonant frequencies in the case of SO coupling for $B=1.2~\mathrm{T}$: $\omega_{10}=\frac{E_1-E_0}{\hbar}$ and $\omega_{20}=\frac{E_2-E_0}{\hbar}$; $E_0, E_1$ and $E_2$ stand for the energies of $|\mathrm{T}_+\rangle$, $|\mathrm{S}\rangle$ and $|\mathrm{T}_0\rangle$ states, respectively.
}
\label{energy_QD2}
\end{center}
\end{figure}

The energy spectrum for the two-electron system in the double quantum dot is displayed in Fig. \ref{energy_QD2}.
We consider the magnetic field of 1.2 T with the system in the $\mathrm{T}_+$ ground state in which the flow of the current across
the two quantum dots is blocked by Pauli exclusion, the hopping of the electron pair into one of the dots is forbidden
by the fact that their spins are parallel. 
We focus on transitions to both states with zero component of the total spin in the $z$ direction of the applied magnetic field, $\mathrm{T}_0$ and $\mathrm{S}$.
In presence of the hyperfine coupling to the nuclear spin field these two states are actually mixed \cite{mixed}, in both the Pauli blockade is lifted \cite{nadji}
and the experimental EDSR spectra resolve both these lines.

\begin{figure}[h!t]
\begin{center}
\includegraphics[width=0.4\textwidth]{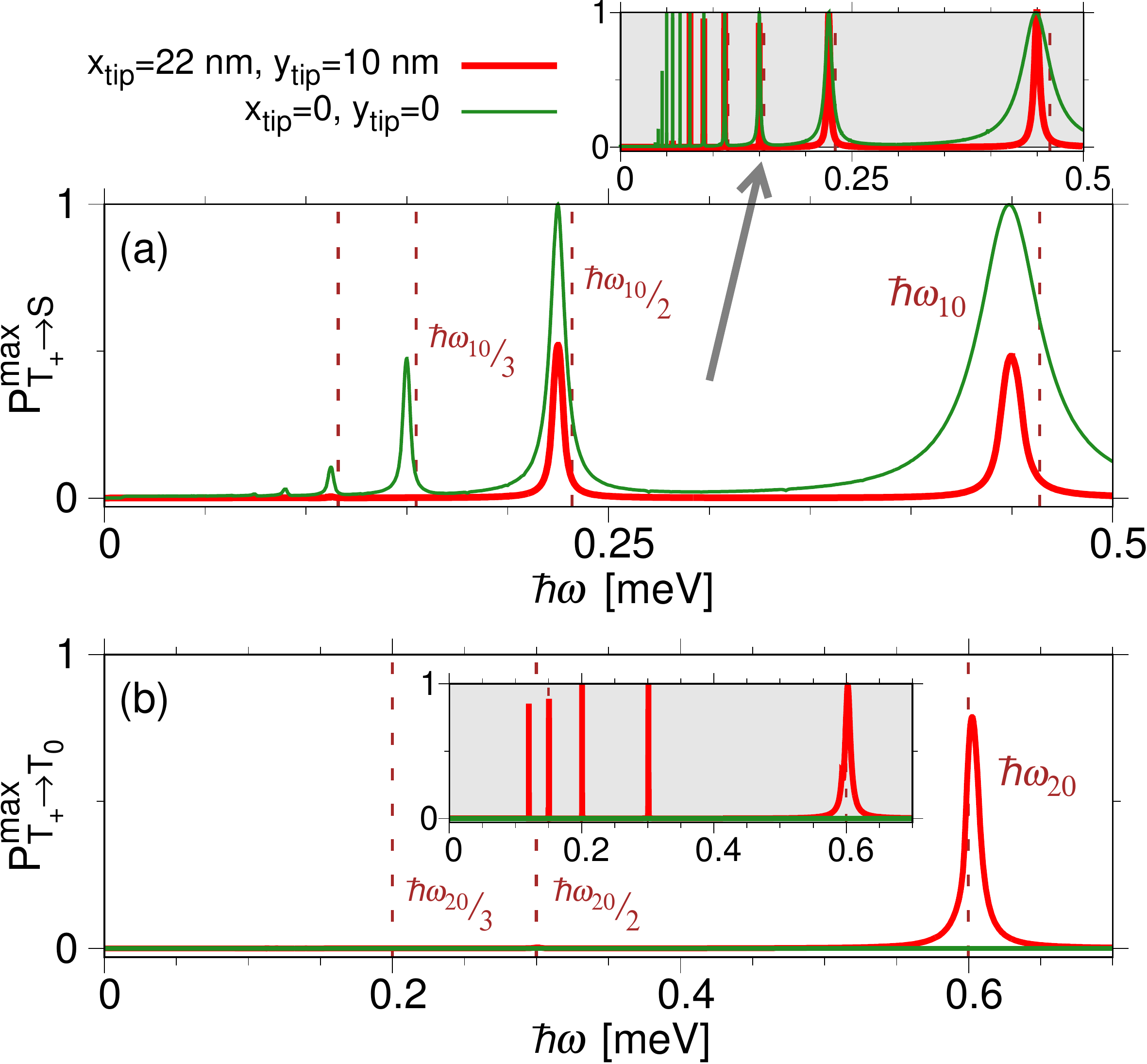}
\caption{Results of the solution of the time-dependent Schr\"odinger equation for the electron pair; $B=1.2~\mathrm{T}$ is assumed. The simulations with the initial state $|\Psi_i\rangle = |\mathrm{T}_+\rangle$ lasted $300~\mathrm{ps}$ ($250~\mathrm{ns}$ for the insets). Maximal probabilities of finding the electron in $|\mathrm{S}\rangle$ (upper panel) and $|\mathrm{T}_0\rangle$ (lower panel) states as functions of the AC frequency for two positions of the tip that are defined in a color legend. Vertical dashed lines indicate a few resonant frequencies: $\hbar\omega_{10}$, $\hbar\omega_{10}/2$, $\hdots$, and $\hbar\omega_{20}$, $\hbar\omega_{20}/2$, $\hdots$. 
}
\label{prob_QD2}
\end{center}
\end{figure}

\begin{figure}[h!t]
\begin{center}
\includegraphics[width=0.4\textwidth]{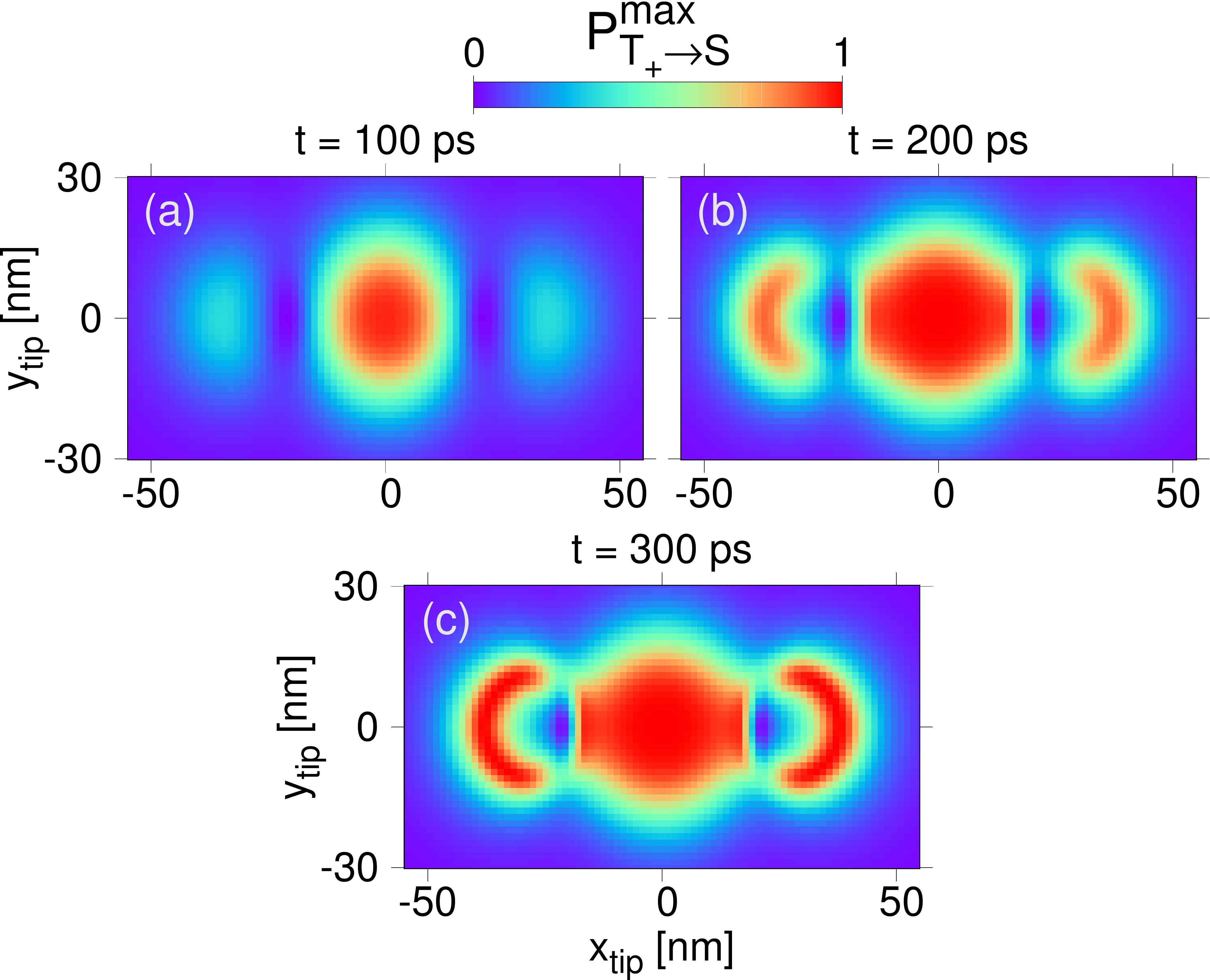}
\caption{Maximal occupation of the singlet state $|\mathrm{S}\rangle$ for simulation lasting (a)~$100~\mathrm{ps}$, (b)~$200~\mathrm{ps}$, (c)~$300~\mathrm{ps}$ as function of the tip position. The frequency $\hbar\omega = 0.4486~\mathrm{meV}$ used in the calculations corresponds to the direct resonance (see Fig.~\ref{prob_QD2}(a)).
}
\label{map10_QD2}
\end{center}
\end{figure}

\begin{figure}[h!t]
\begin{center}
\includegraphics[width=0.4\textwidth]{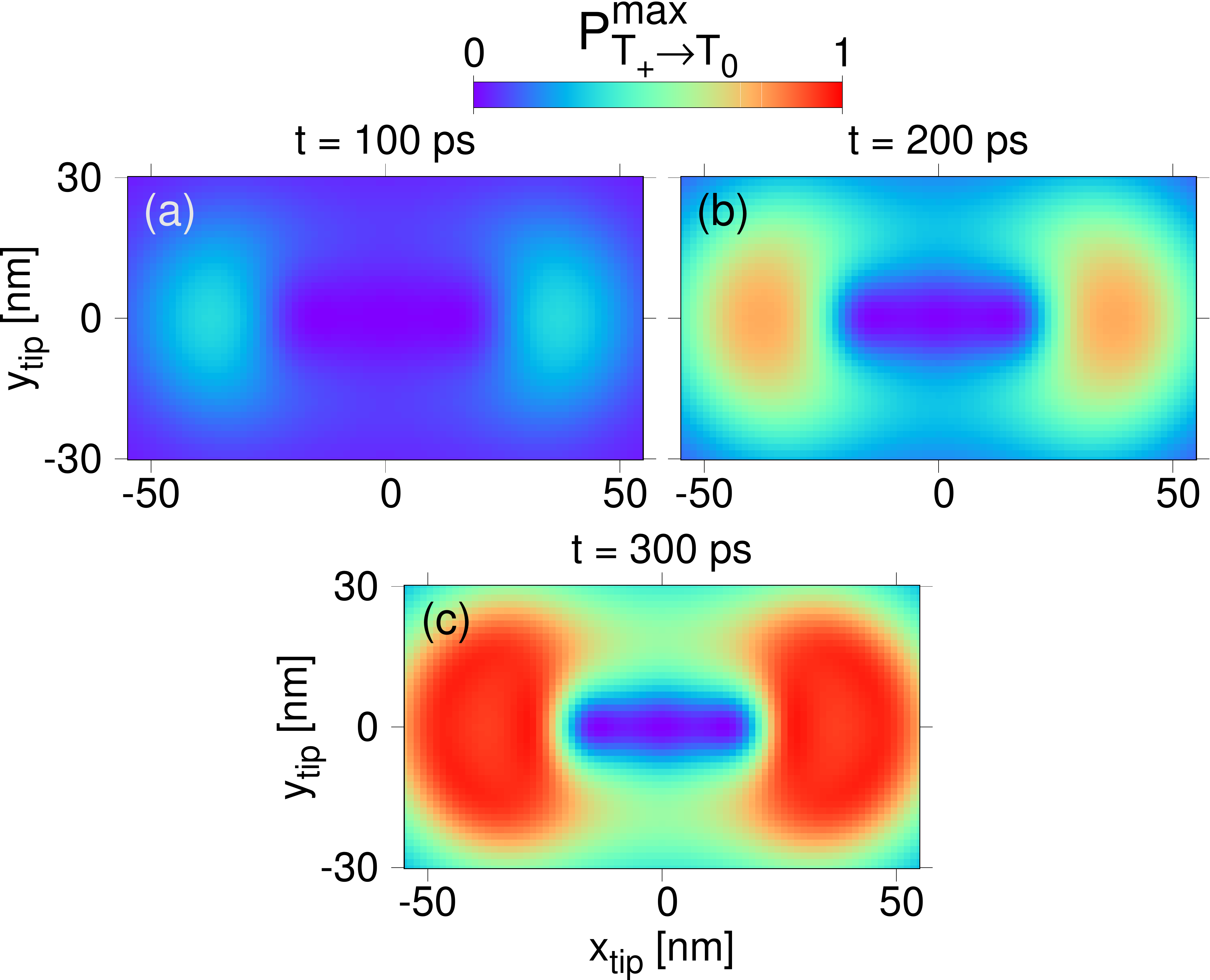}
\caption{Maximal occupation of the triplet state $|\mathrm{T}_0\rangle$ for simulation lasting (a)~$100~\mathrm{ps}$, (b)~$200~\mathrm{ps}$, (c)~$300~\mathrm{ps}$  as function of the tip position. The frequency $\hbar\omega = 0.602~\mathrm{meV}$ used in the calculations corresponds to the direct resonance (see Fig.~\ref{prob_QD2}(b)).
}
\label{map20_QD2}
\end{center}
\end{figure}

In Fig. \ref{prob_QD2}(a) and (b) we present the transition probabilities to the $\mathrm{S}$ and $\mathrm{T}_0$ states, respectively. The dependence of the transition rate 
on the tip position is found as in the single-electron case and for the transition to $\mathrm{T}_0$ state the transition is absent for the tip at the center
of the system. The transition maps as functions of the tip position are given in Figs. \ref{map10_QD2} and \ref{map20_QD2} for $\mathrm{S}$ and $\mathrm{T}_0$ in
the final state, respectively. The maps differ for the tip above the central part of the system, where a maximum (for $\mathrm{S}$ in Fig. \ref{map10_QD2}) or a minimum (for $\mathrm{T}_0$
in Fig. \ref{map20_QD2}) is observed. Local maxima for the tip above the ends of the quantum dots are observed for both the final states, although
for $\mathrm{T}_0$ these maxima are more pronounced. 

\begin{figure}[h!t]
\begin{center}
\includegraphics[width=0.4\textwidth]{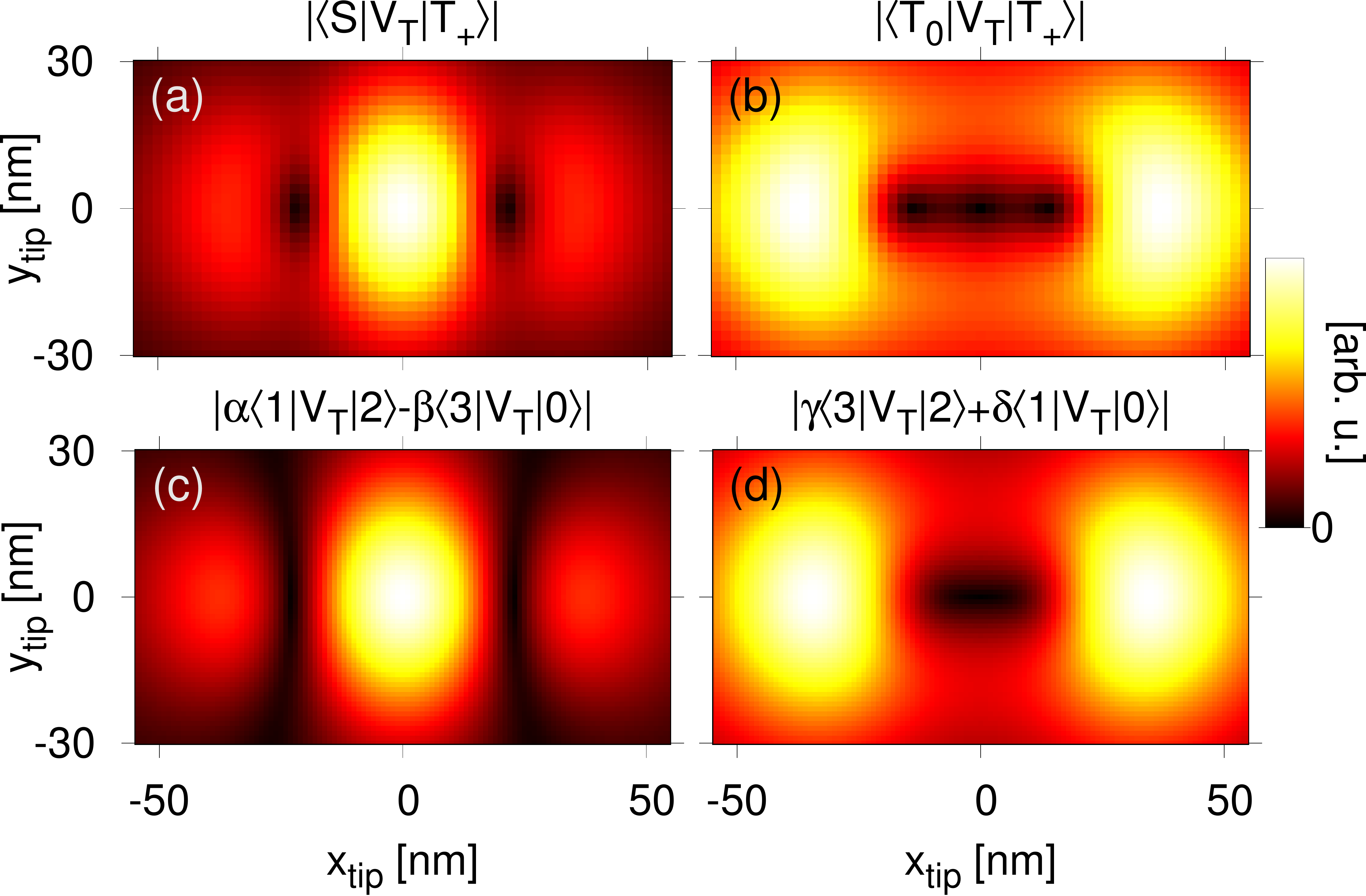}
\caption{The absolute values of (a)~$\langle \mathrm{S}|V_\mathrm{T}|\mathrm{T}_+ \rangle$ and (b)~$\langle \mathrm{T}_0|V_\mathrm{T}|\mathrm{T}_+ \rangle$ matrix elements as functions of the tip position. (c),~(d)~The corresponding approximations of $|\langle \mathrm{S}|V_\mathrm{T}|\mathrm{T}_+ \rangle|$ and $|\langle \mathrm{T}_0|V_\mathrm{T}|\mathrm{T}_+ \rangle|$ obtained with the formulas (\ref{S_Vtip_Tp}), (\ref{T0_Vtip_Tp}).
}
\label{matrix_el_QD2}
\end{center}
\end{figure}

The corresponding matrix elements for the direct transitions are given in Fig. \ref{matrix_el_QD2}(a,b). The general positions of the local extrema of the 
transition probability [Figs. \ref{map10_QD2} and \ref{map20_QD2}] agree with the map of the matrix elements [Fig. \ref{matrix_el_QD2}(a,b)]. The characteristic arc-shaped features of Fig. \ref{map10_QD2}(c) result from the Bloch-Siegert shifts \cite{bs1,bs2,bs3} of the resonant frequency (cf. the dashed line in Fig. \ref{prob_QD2}(a) and the actual position of the direct resonance peak) which vary with the tip position.  The maps in Figs. \ref{map10_QD2} and \ref{map20_QD2} are taken for fixed AC voltage frequency. 

\begin{figure}[h!t]
\begin{center}
\includegraphics[width=0.4\textwidth]{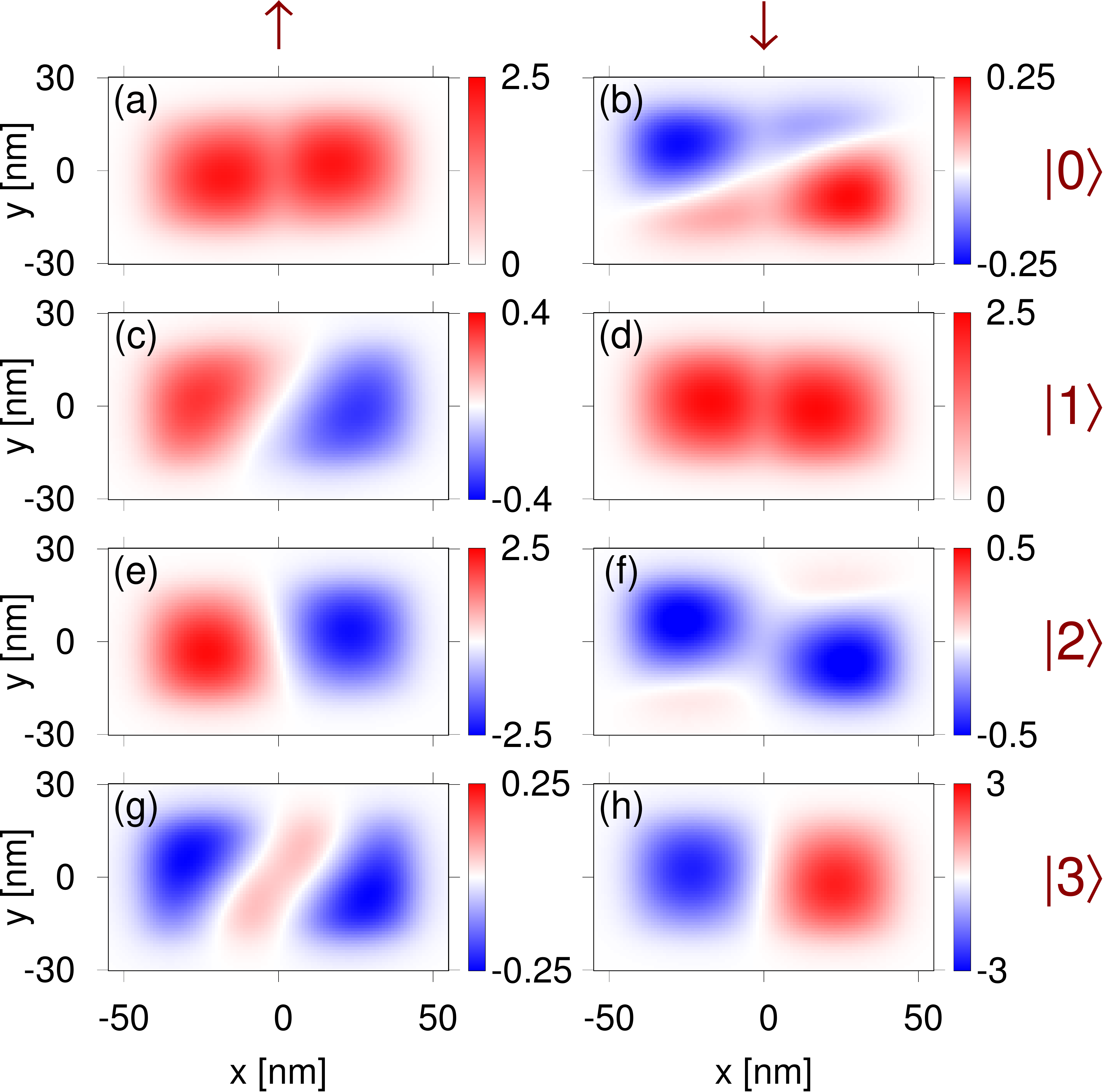}
\caption{Spin up (left column) and spin down (right column) components of the single-electron wave functions for $B=1.2~\mathrm{T}$ in double quantum dot. The color scales give the real part of wave functions in units of $[10^{-2}~\mathrm{nm}^{-1}]$. The ground state is denoted as $|0\rangle$ (cf. first row of the plots), first excited state as $|1\rangle$, etc.
}
\label{wf1e_QD2}
\end{center}
\end{figure}

In order to explain the form of the maps one has to analyze the contributions of the single-electron wave functions to the two-electron states. 
Figure \ref{wf1e_QD2} presents the spin-up and spin-down components of the single-electron states for the double quantum dot.
The wave functions are still eigenstates of $P \sigma_z $ operator, with eigenvalues 1 for the ground state $|0\rangle$ and 
the third excited state $|3\rangle$ and $-1$ for the first and the second excited states ($|1\rangle$~and~$|2\rangle$). 

In Table \ref{Slater_contr} we listed the contributions of the Slater determinants to the two-electron states $\mathrm{S}$, $\mathrm{T}_+$ and $\mathrm{T}_0$.
The two-electron states are eigenstates of the $P(1) \sigma_z(1) P(2) \sigma_z(2)$ with eigenvalues $-1$ for $\mathrm{S}$ and $\mathrm{T}_+$ and $+1$ for $\mathrm{T}_0$.
For the tip located at the center of the system, the symmetry of the potential is unaffected and 
 the transition matrix element $\langle \mathrm{T}_0 | V_\mathrm{T}| \mathrm{T}_+\rangle$ vanishes for the symmetry reasons. Only for the tip moved
off the center the transition can be induced by the AC voltage applied to the tip.  This explains the minimum of the transition maps from $\mathrm{T}_+$ to $\mathrm{T}_0$ 
in the center of the map [Fig. \ref{map20_QD2}]. On the contrary, the transition from $\mathrm{T}_+$ to $\mathrm{S}$ is allowed for the central position of the tip. 

\begin{table}[h!t]
\begin{center}
  \begin{tabular}{r || c  }
~ & $|\mathrm{T}_+\rangle$\\ \hline \hline
$\frac{1}{\sqrt{2}} \left(|02 \rangle -|20 \rangle \right)$ & $0.8806$ \\ \hline
Other Slater determinants & $0.1194$ \\ \hline 
  \end{tabular}
  \vspace*{5pt}
  
    \begin{tabular}{r || c  }
~ & $|\mathrm{S}\rangle$\\ \hline \hline
$\frac{1}{\sqrt{2}} \left(|01 \rangle -|10 \rangle \right)$ & $0.5658$ \\ \hline
$\frac{1}{\sqrt{2}} \left(|23 \rangle -|32 \rangle \right)$ & $0.2792$ \\ \hline
Other Slater determinants & $0.1550$ \\ \hline
  \end{tabular}
  \vspace*{5pt}
  
    \begin{tabular}{r || c  }
~ & $|\mathrm{T}_0\rangle$\\ \hline \hline
$\frac{1}{\sqrt{2}} \left(|03 \rangle -|30 \rangle \right)$ & $0.4925$ \\ \hline
$\frac{1}{\sqrt{2}} \left(|12 \rangle -|21 \rangle \right)$ & $0.4940$ \\ \hline
Other Slater determinants & $0.0135$ \\ \hline
  \end{tabular}
  \caption{Slater determinants forming the basis and their contributions to $|\mathrm{T}_+\rangle$, $|\mathrm{S}\rangle$ and $|\mathrm{T}_0\rangle$ states. The values are found by solving the eigenproblem of two-electron Hamiltonian for a double dot; $B=1.2~\mathrm{T}$.}
  \label{Slater_contr}
\end{center}
\end{table}

For the two-electron wave functions reduced to the principal contributions listed in Table \ref{Slater_contr} the form of the matrix elements
can be expressed in terms of the single-electron matrix elements 
\begin{equation}
 \langle \mathrm{S} |V_\mathrm{T}| \mathrm{T}_+ \rangle \approx \alpha \langle 1 |V_\mathrm{T}| 2\rangle - \beta \langle 3 |V_\mathrm{T}| 0\rangle,
 \label{S_Vtip_Tp}
\end{equation}
and
\begin{equation}
 \langle \mathrm{T_0} |V_\mathrm{T}| \mathrm{T}_+ \rangle \approx \gamma \langle 3 |V_\mathrm{T}| 2\rangle - \delta \langle 1 |V_\mathrm{T}| 0\rangle
 \label{T0_Vtip_Tp},
\end{equation}
where $\alpha$, $\beta$, $\gamma$ and $\delta$ are determined by the diagonalization of the two-electron Hamiltonian matrix in the Slater determinants basis. 
The form of the matrix elements for the approximated two-electron wave functions is given in Fig. \ref{matrix_el_QD2}(c,d) with a good agreement
with the exact maps of Fig. \ref{matrix_el_QD2}(a,b).

\section{Summary and Conclusions}
We have considered resonant spin transitions driven by the AC potential applied to tip of a scanning probe for a 
quantum dot in a semiconductor with spin-orbit coupling. The single-electron states were determined using a mesh of
Gaussian functions and the two-electron states were determined by the configuration-interaction method. The stationary
Hamiltonian eigenstates were used for solution of the system dynamics with AC perturbation introduced by the probe. 
We  demonstrated that the spin-transition rate strongly depends on the tip position and that the transition maps 
correspond closely to transition matrix elements with the tip potential. The latter involve the overlaps of the
minority and majority spin components of the wave functions for  the initial and final states with the tip potential.
This opens an opportunity of experimental probing of the spatial distribution of the spin components for the confined wave functions.
An experiment for the two-electron states in double quantum dot should resolve separate maps for the $\mathrm{T}_+\rightarrow \mathrm{S}$ and $\mathrm{T}_+\rightarrow \mathrm{T}_0$ 
transitions lifting the Pauli blockade of the current flow. The maps bear signatures of the two-electron spin-orbital symmetries
and can be used for identification of the separate transition lines. 

\section*{Acknowledgements}
This work was supported by the National Science Centre according to decision DEC-2015/17/N/ST3/02282  and by PL-Grid Infrastructure.

\end{document}